\documentclass[12pt,preprint]{aastex}
\usepackage{graphicx}
\usepackage{amsmath}
\usepackage{txfonts}
\usepackage{natbib}

\shorttitle{MACHOS}
\shortauthors{Monroy-Rodr\'{\i}guez \& Allen}
\begin{document}

\title{The end of the MACHO era- revisited: new limits on MACHO masses from halo wide binaries}
\author{Miguel A. Monroy-Rodr\'{\i}guez \altaffilmark{1}  \& Christine Allen\altaffilmark{1}}
\affil{Instituto de Astronom\'{\i}a, Universidad Nacional Aut\'onoma de M\'exico, Apdo. Postal 70-264, M\'exico, D.F. 04510, M\'exico; chris@astro.unam.mx}

\begin{abstract}
In order to determine an upper bound for the mass of the massive compact halo objets (MACHOs) we use the halo binaries contained in a recent catalog (Allen \& Monroy-Rodr\'{\i}guez 2013).  To dynamically model their interactions with massive perturbers a Monte Carlo simulation is conducted, using an impulsive approximation method and assuming a galactic halo constituted by massive particles of a characteristic mass. The results of such simulations are compared with several subsamples of our improved catalog of candidate halo wide binaries.  In accordance with Quinn et al. (2009) we also  find our results to be very sensitive to the widest binaries.  However, our larger sample, together with the fact that we can obtain galactic orbits for 150 of our systems, allows a more reliable estimate of the maximum  MACHO mass than that obtained previously.  If we employ the entire sample of 211 candidate halo stars  we obtain an upper limit of $112~M_\sun$. However, using the 150  binaries in our catalog with computed galactic orbits we are able to refine our fitting criteria.  Thus, for the 100 most halo-like binaries we obtain a maximum MACHO mass of $21-68~M_\sun$.  Furthermore, we can estimate the dynamical effects of the galactic disk using binary samples that spend progressively shorter times within the disk.  By extrapolating the limits obtained for our most reliable -albeit smallest- sample we find that as the time spent within the disk tends to zero the upper bound of the MACHO mass tends to less than $5~M_\sun$.  The non-uniform density of the halo has also been taken into account, but the limit obtained, less than $5~M_\sun$, does not differ much from the previous one.   Together with microlensing studies that provide lower limits on the MACHO mass,  our results essentially exclude the existence of such objects in the galactic halo.
\end{abstract}
\keywords{binaries: general -- dark matter -- Galaxy: halo stars: kinematics -- stars: statistics}

\section{Introduction}
\label{sec:intro}
 Disk and halo binaries are relevant to the understanding of processes of star formation and early dynamical evolution.  In particular, the orbital properties of the wide binaries remain unchanged after their formation, except for the effects of their interaction with perturbing masses encountered during their lifetimes, as they travel in the galactic environment.  The widest binaries are quite fragile and easily disrupted by encounters with various perturbers, be they stars, molecular clouds, spiral arms or MACHOs (massive compact halo objects).  For this reason, they can be used as probes to establish the properties of such perturbers.  The use of wide binaries as sensors of the dynamical effects of perturbers in the galactic halo has been the subject of several studies (Yoo et al. 2004, Quinn et al 2009).
	An interesting application of halo wide binaries was proposed by Yoo et al. (2004) who used 90 halo binaries from the catalog of Chanam\'e \& Gould (2004). To try to detect the signature of disruptive effects of MACHOs in their widest binaries.  They were able to constrain the masses of such perturbers to  $M < 43~M_\sun$, practically excluding MACHOs from the galactic halo, since a lower limit of about $30~M_\sun$ is found by other studies (Tisserand et al. 2007, Wyrzykowski et al. 2008).  However, as was  shown a few years later by Quinn et al. (2009) this result depends critically on the widest binaries of their sample, as well as on the observed distribution of the angular separations which, according to Chanam\'e \& Gould, shows no discernible cutoff between $3.5\arcsec$ and $900\arcsec$.  Quinn et al. (2009) obtained radial velocities for both components of four of the widest binaries in the Chanam\'e \& Gould catalog and found concordant values for three of them, thus establishing their physical nature.  The fourth binary turned out to be optical, resulting from the random association of two unrelated stars. Quinn et al. excluded this spurious pair from their study.  They assumed a power law distribution for the initial binary projected separation and varied the exponent over a wide range.   They evolved the distributions and compared their results with the distribution of the observed projected separations, attempting a best fit of the models to the observed distribution.
Quinn et al. found a limit for the masses of the MACHOs of $M < 500~M_\sun$, much less stringent than that previously obtained.  They concluded rather pessimistically that the currently available wide binary sample is too small to place meaningful constraints on the MACHO masses.  They also pointed out that the density of perturbers encountered by the binaries along their galactic orbits is variable, and that this variation should be taken into account when calculating constraints on MACHO masses.

	Motivated by these concerns, we have constructed a catalog of 211 candidate halo wide binaries (Allen \& Monroy-Rodr\'{\i}guez 2013), and computed galactic orbits for 150 such binaries.  The orbits allowed us to establish the fraction of its lifetime each binary spends within the galactic disk and thus to estimate the dynamical effects of the passage of the binary through the disk.  The orbits also allow us to estimate the effects of the variable halo density encountered by the binaries along their galactic trajectories.  We stress that for a total of 9 of the widest binaries concordant radial velocities for both components were found in the literature, thus confirming their physical association.
The radial velocities of both components of these pairs were found to be equal within the published observational uncertainties.  A more sophisticated analysis was not attempted because the random association of two unrelated stars will most probably result in grossly discrepant radial velocities.  Inspection of Table 1 in Quinn et al. (2009) shows that, indeed, the radial velocities of their three physical pairs are equal within the stated uncertainties, whereas the components of the non-physical pair have radial velocities differing by almost 80 km/s.

\section{The dynamical model}
To model the dynamical evolution of our sample of halo wide binaries we follow closely the procedure of Yoo et al. (2004).  In this approach, the impulse approximation is used, and effects of large-scale galactic tides and molecular clouds are neglected. Yoo et al. (2004) give a detailed description of the impulse approximation. Dissolved binaries are eliminated from the modelled population. The galactic mass distribution is assumed to be constant over time. Thus, the effects of major mergers are ignored. Such mergers would possibly be quite destructive of fragile systems such as wide binaries. On the other hand, they could also be a source of fresh wide binaries acquired by our Galaxy (Allen et al. 2006). The dynamical and merger history of our galaxy is at present too poorly known to take into account in our model.

We initially assume a constant halo density, and attribute to the MACHOs the total local halo density, taken as $\rho =  0.007~M_\sun/{\rm pc}^3$.  This value is taken from Allen \& Santill\'an (1991) mass model, for consistency with the computed galactic orbits of the wide binaries. It is quite similar to the values used by Yoo et al. (2004) and Quinn et al (2009). It also agrees with the value obtained by Widrow et al. (2008). We then evaluate the effects of the 100 closest encounters in the tidal regime, and compute the cumulative effects of weak encounters in the Coulomb regime.

We start the Monte Carlo simulations by drawing initial semiaxes uniformly from a power law distribution with an arbitrary exponent, and with separations between 10 AU and 300 000 AU The mass of each binary component is taken as $0.5~M_\sun$, and the velocity distribution of the perturbers is assumed to be isotropic with a dispersion of 200 km/s in each component, appropriate for halo objects.  In each model simulation we let  100 000 binary systems evolve for 10 Gy.  For each model (with a fixed MACHO mass $M$ and a fixed local halo density), we construct a scattering matrix to simultaneously investigate large sets of  initial power law distributions.  This scattering matrix gives the probability that a binary with initial semiaxis $a$  will have a semiaxis $a^\prime$ after 10 Gyr of evolution.  For each model we generate a virtual binary sample with the same number of binaries as the observed sample, that is, such that the number of binaries with major semiaxes between $a_{{\rm min}}$ and $a_{{\rm max}}$  is equal to the observed number of binaries within these limits.  These limits will be different for the different subsamples used to estimate the maximum MACHO mass.  The virtual binary samples resulting from the evolved model are then compared with the observed samples of binaries. For the comparison with the binned observational data, each bin of virtual binaries is taken as the centroid of the corresponding group of model binaries, comprising about 10,000 binaries. For comparison with the unbinned observational data each virtual binary is constructed as the centroid of more than 500 evolved binaries (for the comparison with the 100 most halo-like binaries), and more than 2600 (for comparing with the 25 most halo-like binaries). In all cases, therefore, each virtual binary represents a sufficiently large number of evolved model binaries, and hence sampling errors are negligible.

	The goodness of fit is evaluated by minimizing the dispersion, $\sigma$ defined as:

\begin{equation}
\sigma = \frac{\sqrt{\sum_{i} \left(\langle a\rangle_{{\rm obs}} - \langle a\rangle_{{\rm virtual}}\right)^2}}{N} ,
\nonumber\\
\end{equation}

\noindent where $\langle a\rangle_{{\rm obs}}, \langle a\rangle_{{\rm virtual}}$ are the expected major semiaxes of the observed and the virtual binary, respectively, and $N$ is the number of binaries in the sample. For the purpose of comparing our results with previous ones, we also computed the $\sigma$ from a maximum likelihood estimate, adhering to the definition of $\sigma$ used by Quinn et al (2008).  We  stress that the distributions of semiaxes of the simulated and the observed binary samples are compared directly. Previous work (Yoo \& Chaname, Quinn et al. 2009) fitted the observed projected separations. Since distances are available for all our binaries, we prefer to compare directly the resulting model distribution of major semiaxes with the observed one, calculated from the angular separations and the individual distances.  Projection effects are taken into account by means of the widely used statistical formula (Couteau 1960):

\begin{equation}
\langle a\arcsec\rangle = 1.40 s\arcsec . \nonumber\\
\end{equation}

This formula was derived theoretically, and includes the eccentricity distribution of the binaries, as well as projection effects. Here, $\langle a\rangle$ represents the expected value of the major semiaxis $a$ and $s$ the angular separation. A more recent empirical study (Bartkevicius 2008) compares the angular separations with the major semiaxes for systems from the Catalog of Orbits of Visual Binaries (Hartkopf \& Mason 2007).  The results obtained vary somewhat according to the data used for the separations $s$, either from ``first''  or  ``last'' observations in the WDS (Mason et al. 2001) , or from the orbit catalog ephemerides, the latter deemed to provide the most reliable value, namely:

\begin{equation}
\log\langle a\rangle - \log s = 0.112. \nonumber\\
\end{equation}

\noindent In spite of the rather large uncertainties of the latter values, we shall use both estimates to compute the maximum MACHO masses.

The individual distances to the binaries used to convert separations in arcsec into separations in AU and then into expected major semiaxes have, of course observational uncertainties.  Since the groups of binaries we will use to determine maximum MACHO masses contain binaries from different sources, we can only provide an estimated average uncertainty for the distances, trying to err on the conservative side.  In our catalog, we adopted the distances  given in the Hipparcos catalog when their  parallax errors were  less than 15\%,  Otherwise we adopted, in order of preference, the Str\"omgren photometric distances given in the lists of Nissen \& Schuster (1991) , those obtained with the polynomial of Chanam\'e \& Gould, using their photometry, or the spectroscopically derived distances of Ryan (1992) and Zapatero-Osorio \& Martin (2004).  Most of our binaries should have distance errors of less than about 15\%.

To obtain the values of the power-law exponent and the maximum MACHO mass compatible with the observations we proceed as follows.
We start by assuming the mass of the perturbers to be zero, and multiply the scattering matrix by power laws, letting the exponent vary between $-1$ and $-2$ (in increments of $0.01$), until the best fit of the virtual binary sample to the observed distribution is found, where by ``best fit'' we mean the fit that minimizes the dispersion $\sigma$, as defined  above.  This fit is characterized by a dispersion $\sigma = \sigma_0$.  Then we repeat the procedure, increasing the perturber mass in increments of $1~M_\odot$, until a fit is obtained with a $\sigma = 2 \sigma_0$, which we take as still acceptable.  This yields a more conservative estimate of the maximum MACHO mass, but  still compatible (within $2 \sigma_0$) with the observations.

Our implementation of the dynamical model was tested by applying it to  the binary samples studied by Yoo et al. (2004) and Quinn et al. (2009).

Figure 1 shows the results of applying our dynamical model to the binaries of Yoo et al. Qualitatively, it is seen that perturber masses of $1000~M_\odot$ and $100~M_\odot$ appear to be excessive, while perturber masses of $10~M_\odot$ are insufficient to account for the observed distribution. This figure shows results very similar to those of previous studies (see e.g. Fig. 5 in Yoo et al.).

Figure 1  also shows that the  horizontal line (exponent -1) represents the initial -unevolved- distribution which has to evolve farthest in order to fit the observed distribution, that is, that which needs the largest mass of perturbers to conform with observations.  In this sense, taking an initial distribution with a -1 exponent yields the most conservative estimate for the perturber mass, i.e. the largest MACHO mass.

Figure 2 shows the exclusion contours we obtain by applying our model to the Yoo et al. and Quinn et al. samples. Since we are comparing semiaxes instead of projected angular separations an exact match is not to be expected. Nonetheless our results closely agree with previous ones, specially in the region of large halo densities and small perturber masses, which is the interesting one.

The range of exponents considered for the initial power law distribution of our model binaries amply covers the observationally found ones.  So, for example, Chanam\'e \& Gould (2004) find $-1.67$ and $-1.55$ for their disk and halo binaries, repectively.    Sesar et al. (2008)  find  $-1$;  Lepine \& Bongiorno (2007) obtain $-1$;  Poveda et al. (1994, 1997, 2003, 2004)  find -1 but up to different limiting semiaxes according tto the age of the binaries;  finally, Allen et al. (2000) find $-1$ for their  high-velocity metal-poor binaries, but also only up to different limiting semiaxes, according to the time spent by the binaries within the galactic disk.

Furthermore, as pointed out by  also by Quinn et al., there are both theoretical and observational reasons to justify the assumption of an exponent $-1$ for the unevolved distribution. Indeed, Valtonen (1997) showed Oepik's distribution to correspond to dynamical equilibrium, and it has been found to hold for many different samples of wide binaries of different ages and provenances (Poveda et al. 2006, Sesar et al. 2008, Lepine \& Bongiorno 2007,  Allen et al.  2000).  In the accompanying paper (Allen \& Monroy-Rodr\'{\i}guez 2013) we show that it holds for halo binaries up to different limiting semiaxes,  according to the time the binaries spend within the galactic disk.  For the most halo-like binaries, it holds up to expected semiaxes of 63,000 AU.

	As pointed out by Quinn et al. the results are dependent on the definition of $\sigma$. We adhere to the definition of Quinn et al. to compare our results with theirs. However, we found that the maximum likelihood method does not give as good a resolution for the MACHO mass as that obtained using the ordinary dispersion $\sigma$ as defined above, or using the KS indicator for the unbinned distributions, as we describe below.

	The best fit between the virtual binary distribution and the observed one fixes the value of $M$, the mass of the perturbers. As explained above, the most conservative estimate (largest MACHO mass) is obtained for an exponent of $-1$, which corresponds to the  Oepik distribution.

As an illustration, Figure 3 shows  the results of the numerical simulation compared to the observed groups of the 100, 50 and 25 most halo-like binaries from our catalog.  The frequency distribution of expected semiaxes is plotted in the vertical axis.  The large dots indicate the observed distribution, with error bars corresponding to sampling errors.  The horizontal line corresponds to an unevolved binary population with a power law exponent of $-1$. The dashed lines shows the best fit we obtain for the observed distribution, using two power laws for the unperturbed and perturbed part, respectively (see below).  The figure shows qualitatively how, depending on the observed sample, the admissible perturber masses range from about $50~M_\odot$ to less than $10 ~M_\sun$.

To construct Figure 3, we started by fitting each observed sample with a power law of exponent $\alpha = -1$ up to the value of the semiaxis where dissolution effects become noticeable.  As shown in the companion paper (Allen \& Monroy 2014) the Kolmogorov-Smirnov (KS) test shows this fit to be valid up to different values of the semiaxes. We refer to this region as the unperturbed region. For the perturbed region (larger values of the semiaxes) we use a steeper power-law fit. The KS test shows (see e.g. Figure 5) that using two power-law fits allows an excellent representation of the observed samples for the entire range of semiaxes.

The virtual binary samples obtained from the evolved model binaries are then compared with these fits.  Figure 4 shows an example of the comparison with the samples of the 100, 50 and 25 most halo-like stars.  The large dots represent the observed sample. The full squares were obtained by varying the perturber mass in intervals of $1~M_\odot$, until the standard deviation $\sigma$ attained its minimum value;  let this value be called $\sigma_0$.  We then continue increasing the perturber mass until a $\sigma = 2\sigma_0$ is reached, which we consider still acceptable and which is represented in Figure 4 by the empty triangles.  This gives a conservative estimate the maximum permissible perturber mass. The right-hand panels show the result of applying the KS test to the fits. The KS indicator, $Q$, shows an excellent fit (large dots) to the full squares (corresponding to $\sigma_0$), and an acceptable fit (small dots) to the triangles (corresponding to $2\sigma_0$) for all but the largest semiaxes.

To further refine our results we tried an alternative method. Figure 5 shows the result of applying the KS test to different observed samples, using two power-law fits for the unperturbed and perturbed regions, respectively. The figure shows an excellent agreement ($Q=1$, as shown in the right hand scale) for the entire observed range of semiaxes.

We now compare the cumulative distribution of semiaxes of the virtual binaries obtained from the dynamical model to these fits, using again the KS test as criterion of goodness of fit and constructing the virtual binary samples as described above. Figure 6 illustrates the results we obtain for the 100, 50, and 25 most halo-like binaries. The full diamonds denote the observed samples, the crosses the best fits (which fix $\sigma_0$), and the empty triangles the still permissible fits (corresponding to $2\sigma_0$). The goodness of these fits, as shown by the KS-indicator $Q$ is displayed in the right hand scale of the panels.
This methods avoids binning the data, and thus allows a better resolution. Even for the comparison with the 100 most halo-like stars, the statistical errors of the modeled binaries remain negligible, since each virtual binary is constructed as the centroid of more than 500 evolved model binaries.

\section{Application to the improved catalog of candidate halo binaries}
As shown above for three samples, we are now ready to apply our dynamical model to different subsets of binaries extracted from our catalog (Allen \& Monroy 2013).  For this study we consider mainly the 150 binaries with computed galactic orbits. The galactocentric orbits were integrated using the standard 1991 Allen \& Santillan (axysimmetric)  galactic potential.  Since very few orbits approach the zone of influence of the bar, an axisymmetric potential was found to be adequate.  The orbits were integrated backwards in time for 10 Gyr.  The relative energy error amounted to less than $10^{-7}$ at the end of the integration.
 To establish membership to the halo we calculate the time the binary spends within the galactic disk ($z= \pm 500$~pc).  In Table 1 we display our main results.  The  different samples of binaries listed in Column 1  are ordered by the time they spend in the disk.  Thus, the first entry corresponds to the 25 most halo-like binaries, which spend an average of only $0.08$ of their lifetimes in the disk.  The second column lists the fraction of their lifetimes the binaries spend in the disk.  Column 3 displays the upper limits obtained for the binned data for each group of binaries, with the fits performed to $2\sigma$. Column 4 shows the limits obtained taking into account, on average,  the variable halo density encountered by the binaries along their galactic trajectories. These two columns are labeled with the subscript b, to denote that the results were obtained from the binned data. Column 5 shows the MACHO mass limits we obtain using our alternative fitting method (KS) which avoids binning the data. Column 6 lists the corresponding results for a variable halo density. We shall further discuss these entries in Sections 4 and 5. Columns 5 and 6 are labelled KS, to denote that results were obtained using the KS method.

\section{Estimate of the dynamical effect of the disk}
We still have to take into account the dynamical interactions that the binaries experience during their passage through the galactic disk.  These effects have not been considered  in previous studies (Yoo et al. 2004; Quinn et al. 2009).  At this stage, we shall not attempt a full dynamical modeling, but instead roughly estimate  the effects of the disk passages by studying groups of binaries that spend progressively smaller fractions of their lifetimes within the disk. These fractions were obtained from the galactic orbits computed with the Allen \& Santill\'an (1991) mass model. The result of this estimate is shown in Figure 7  where we display the MACHO masses obtained by applying the dynamical model to the  groups of binaries listed in Table 1.  The error bars in Fig. 7 represent an estimate of the effect of the distance errors of the observed samples. Since we order our binaries by the time they spend within the disk, only average values of the distance uncertainties can be given for each group.  These averages translate into average uncertainties in the expected major semiaxes.  We re-computed the dynamical models to obtain best fits with  probable errors of 20\% added to and subtracted from the expected semiaxes.  We used these results to determine the error bars shown in Figure 7.  On average the maximum MACHO masses vary by about 20\%, somewhat more than they do by adopting two alternative formulae for transforming the observed linear separations into expected major semiaxes. These results are displayed in Table 2.
Taking the last point in the graph, we find a maximum MACHO mass of less than  $5~M_\sun$.

We remark that the smallest group, the 25 most halo-like binaries includes 6 very wide systems with concordant radial velocities, as reported in the literature.  Therefore, their physical association can be considered as established.  Five of these binaries are among the widest known, and thus particularly suitable for determining upper limits to the masses of the perturbers.  Since the 25 most halo-like binaries form a very interesting group for further studies (especially, for determining the radial velocities of both components, and thus confirming -or not- their physical nature), we list them in Table 3, along with their main characteristics.

Columns 1 and 2 contain the designation of the primary and secondary components, respectively,  Column 3 lists our adopted distance.  Columns 4 and 5 display the absolute visual magnitudes of the primary and secondary, respectively.  Column 6 lists the projected separation in arcsec, Column 7 the expected value of the major semiaxis in AU.  Column 8 displays the peculiar velocity of the binary, and Columns  9, 10  and 11 the main characteristics of its galactocentric orbit, namely, the maximum distance from the galaxtic center, the maximum distance from the galactic plane, and the eccentricity, respectively.  In the last column we list the fraction of its lifetime spent by the binary in the galactic disk, that is, within $z = \pm 500$~pc.

\section{Estimate of the effect of considering a non-uniform halo density}
As was already pointed out by Quinn et al. the density of the halo cannot be assumed to be uniform.  This is particularly relevant for galactic orbits that reach large apogalactic distances or large distances from the plane.   To obtain a rough estimate of the effect of this non-uniform density on the computed mass limits, we calculated for each binary a time-averaged halo density over its galactic orbit, using the halo mass distribution of Allen \& Santill\'an (1991). Then, we averaged over the binaries in each studied group.  The average densities turned out not to be very different from the density at the solar distance.  This is due to the fact that, since the sample of binaries is restricted to nearby stars,  orbits that reach large apogalactic distances also penetrate close to the galactic center, where they encounter large densities.  Only a few binaries reaching large $z-$distances encounter  considerably smaller average halo densities.  As an aside, we point out that our orbit for NLTT 10536/10548 differs significantly from that plotted by Quinn et al.  Our orbit reaches a $z_{\rm max}$ of only $5.2$~kpc, instead of the almost 40 kpc in Quinn et al. This may partially account for the differences in the average halo densities found by us, which are much less significant than those obtained by the former authors.
	The result of estimating the effects of a variable halo density is shown in Figure 8, for the groups of binaries of Table 1.  The figure shows the mass limits obtained as a function of the fraction of time spent in the disk for each group, for the averaged densities as described above and can be compared with Figure 7. The fits to obtain the limiting masses were performed using the unbinned data sets. From the last point in the graphs it is seen that the limiting masses obtained are less than $5~M_\sun$.   If we extrapolate to zero time spent in the disk we see that the maximum MACHO decreases to about $3~M_\sun$.

	If instead of using the Couteau (1960) formula for transforming observed separations into expected major semiaxes we use the expression derived by Bartkevicius (2008) we obtain the results shown in Table 4.  The table shows that the derived limits increase somewhat, but this increase (on average about 5\%) does not significantly alter the main conclusions of this study.

Finally Fig. 8 shows the exclusion contours resulting from the present study and obtained using the prescription of Quinn et al. (2009). Our results, taken together with the microlensing data, practically exclude the presence of MACHOS in the galactic halo.

\section{Summary and conclusions}

From a catalog of 251 candidate halo wide binaries we have extracted different subgroups in order to determine upper bounds to the MACHO masses.  For a group of 150 binaries it was possible to compute galactic orbits and thus to evaluate the time they spend within the galactic disk.  We developed and validated a dynamical model for the evolution of wide halo binaries subject to perturbations by MACHOs.  This model was applied to different subsamples from the catalog , thus allowing us to obtain as upper limits for the masses of MACHOs the following values:

From 211 systems likely to be halo binaries: $112~M_\sun$.

From 150 halo binaries with computed galactic orbits: $85~M_\sun$.

From 100 binaries that spend the  smallest times within the  disk (on average, half their 			 lifetimes): $21-68~M_\sun$.

	From the same 100 binaries, but taking into account the non-uniform halo density: $28-78~M_\sun$.

	From the 25 most halo like binaries (those that spend on average 0.08 of their lifetimes within 			 the disk): $3-12~M_\sun$.

MACHOs  have been a strong candidate for the dark matter that appears to dominate the mass of galaxies, at least at large radii. Microlensing experiments (Tisserand et al. 2007, Wyrzykowski et al. 2008) exclude baryonic MACHOs with masses in the range $10^{-7}$ to $30~M_\sun$ as dominant constituents of the dark matter halo.  Other studies (Lacey \& Ostriker 1985, Rujula et al. 1992) have ruled out MACHO masses larger than about  $10^6~M_\sun$.  Wide binaries, as studied by Yoo et al. restricted the range of possible MACHO masses to $30-45~M_\sun$, but their results were shown by Quinn to critically depend on one of their widest binaries, which turned out to be spurious.  Excluding this binary, Quinn et al. found a less stringent upper limit of  about $500~M_\sun$.

	In the present study we have used an expanded catalog of candidate halo wide binaries along with their galactic orbits to revise the limits for the MACHO masses.  From different subsamples of binaries we obtain maximum MACHO masses ranging from $112~M_\sun$ all the way down to less than $10~M_\sun$,  The most stringent limits are obtained from the most halo-like sample of 25 wide binaries, a sample particularly suited for further studies.   These results, once again, all but exclude the existence of MACHOs in the galactic halo.

\acknowledgments

We thank A. Poveda for fruitful discussions. MAMR is grateful to UNAM-DGEP for a graduate fellowship. We also thank two anonymous referees for their critique of the paper, which resulted in many improvements.

\clearpage

\begin{figure}[!t]
\includegraphics[width=\columnwidth]{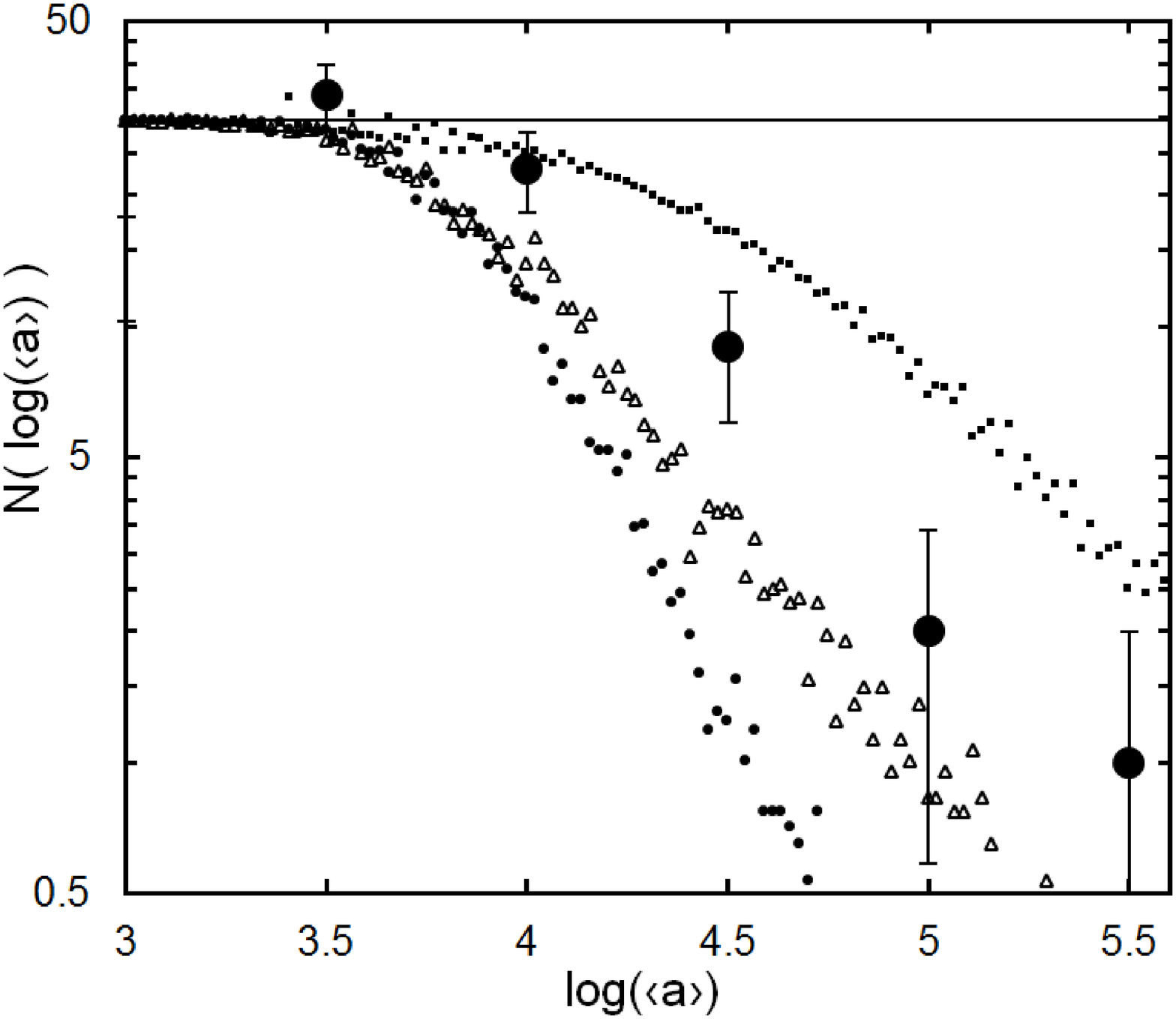}
\caption{Evolved distribution of semiaxes for various perurber masses.  The model was applied to the Quinn et al. binaries.  Dots correspond to perturber masses of $1000~M_\sun$, triangles to $100~M_\sun$, small squares to $10~M_\sun$. Each symbol represents 500 modelled binaries. The graph shows  results very similar to those of Yoo et al. and Quinn et al. Perturber masses of $1000~M_\sun$ appear to be too large, while masses below $10~M_\sun$ seem too small to reproduce the observed distribution (large dots).  The bars in the observed distribution correspond to sampling errors}
\label{fig:simple}
\end{figure}

\begin{figure}[!t]
\includegraphics[width=\columnwidth]{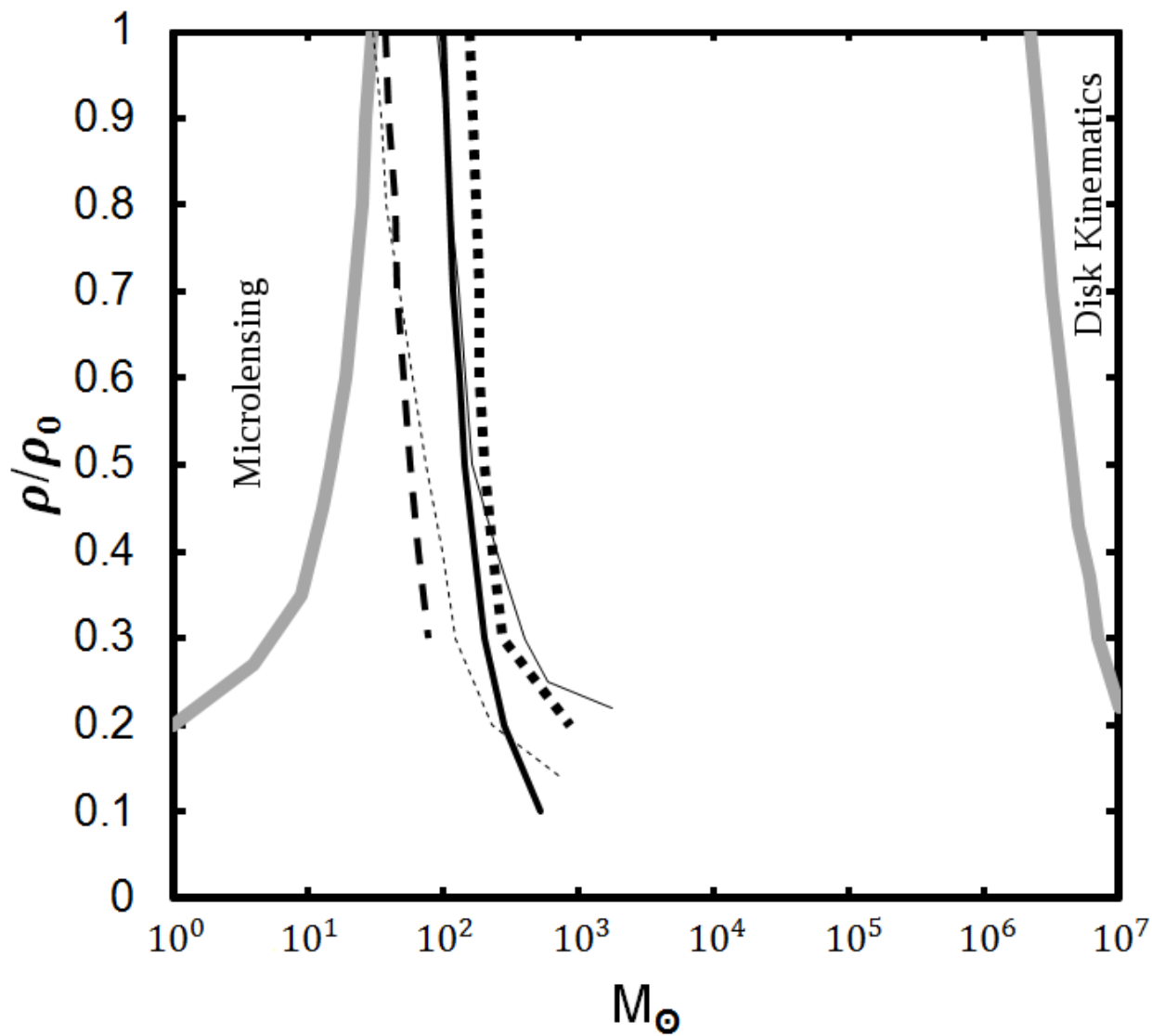}
\caption{Exclusion contour plot. The thick grey lines denote the limits from microlensing experiments and galactic disk stability. The thin dashed line is the result of Yoo et al. for 90 binaries. The thick dashed line is our result for the same sample, using a their method. Both results agree reasonably well. The thin full line is the result of Quinn et al. for the 90 original Yoo et al. binaries.  The thick full line is our result for the same sample, using a fit to $2\sigma$.  Here again, the results agree well.  Finally, the thick  dotted line refers to the result of Quinn for 89 binaries, omitting one spurious pair.  The concordance of previous results with our method applied to the same samples of binaries appears satisfactory, especially in the region of large halo densities and small perturber masses, which is the interesting region. Previous work dealt with projected angular separations.  We deal with semiaxes, and therefore the results cannot be expected to be identical.}
\label{fig:simple}
\end{figure}

\begin{figure}[!t]
\includegraphics[height=18cm]{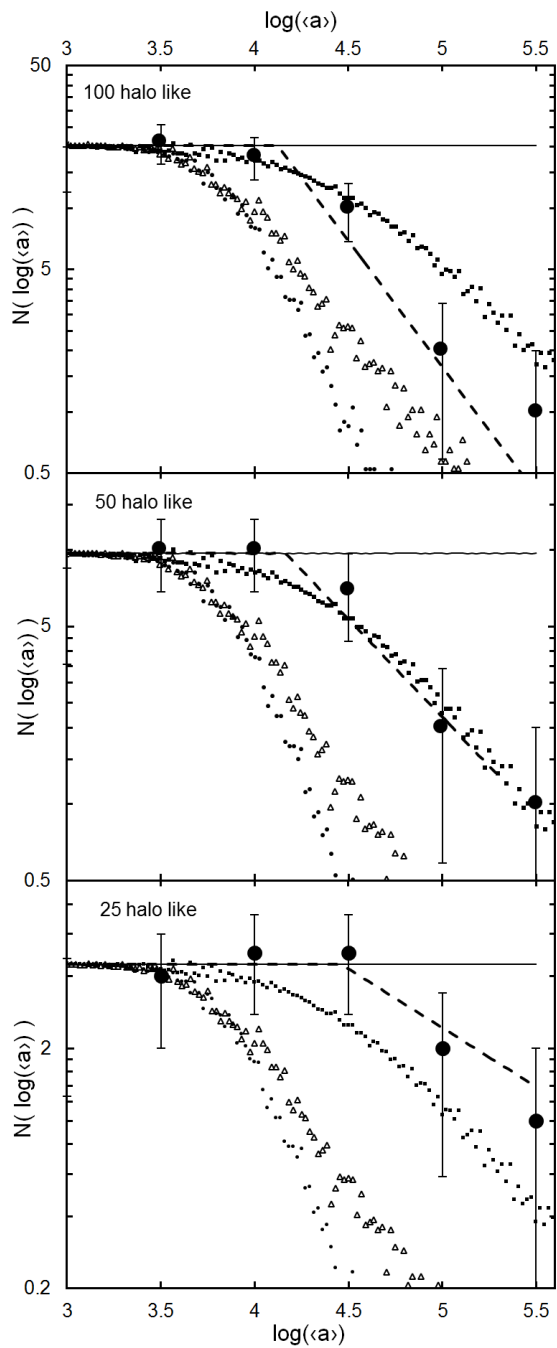}%width=\columnwidth]
\caption{The evolved distribution of semiaxes for different perturber masses, compared with three samples of binaries from our catalog. The straight horizontal line shows the initial distribution.  The large dots with error bars denote the observed data.  The dashed lines show the two power-line fits to the observed samples, namely an Oepik law fit to the unevolved region and a steeper power law for the evolved region (see text). Filled squares correspond to perturber masses of $10~M_\sun$, triangles to $100~M_\sun$ and dots to $1000~M_\sun$.  Depending on the sample considered, perturber masses of several hundreds $M_\sun$, about $10~M_\sun$ and less than $10~M_\sun$ appear to fit respectively the 100, 50 and 25 most halo-like  binaries.}
\label{fig:simple}
\end{figure}

\begin{figure}[!t]
\includegraphics[height=18cm]{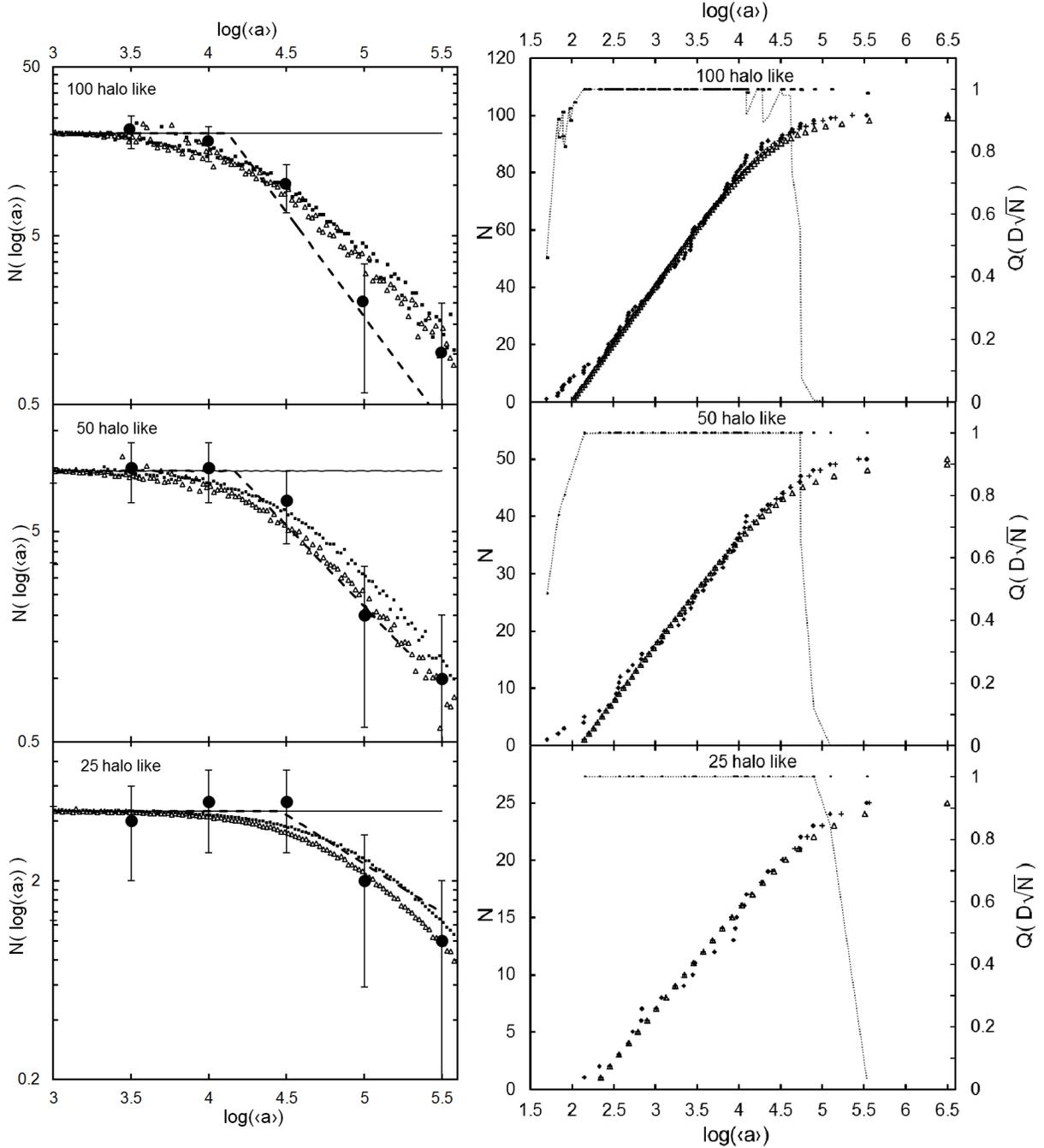}%width=\columnwidth]
\caption{Fits of the evolved model distribution to different samples of binaries.  The squares show the best fit, which defines $\sigma_0$ (see text), and corresponds to perturber masses of 16, 7 and $2~M_\sun$ for the 100, 50 and 25 most halo-like stars, respectively.  The triangles correspond to fits to $2\sigma_0$, which we still consider acceptable, and give more conservative estimates for the perturber mass:  21, 11 and $3~M_\sun$ for the 100, 50 and 25 most halo-like stars, respectively. The right-hand panels display the results of applying the Kolmogorov-Smirnov test to these fits. The large dots correspond to the squares ($\sigma_0$) and show that the fit is excellent ($Q=1$) for the entire range of semiaxes. The small dots correspond to the triangles ($2\sigma_0$) and show that the fit is still acceptable for all but the largest semiaxes.}
\label{fig:simple}
\end{figure}

\begin{figure}[!t]
\includegraphics[height=18cm]{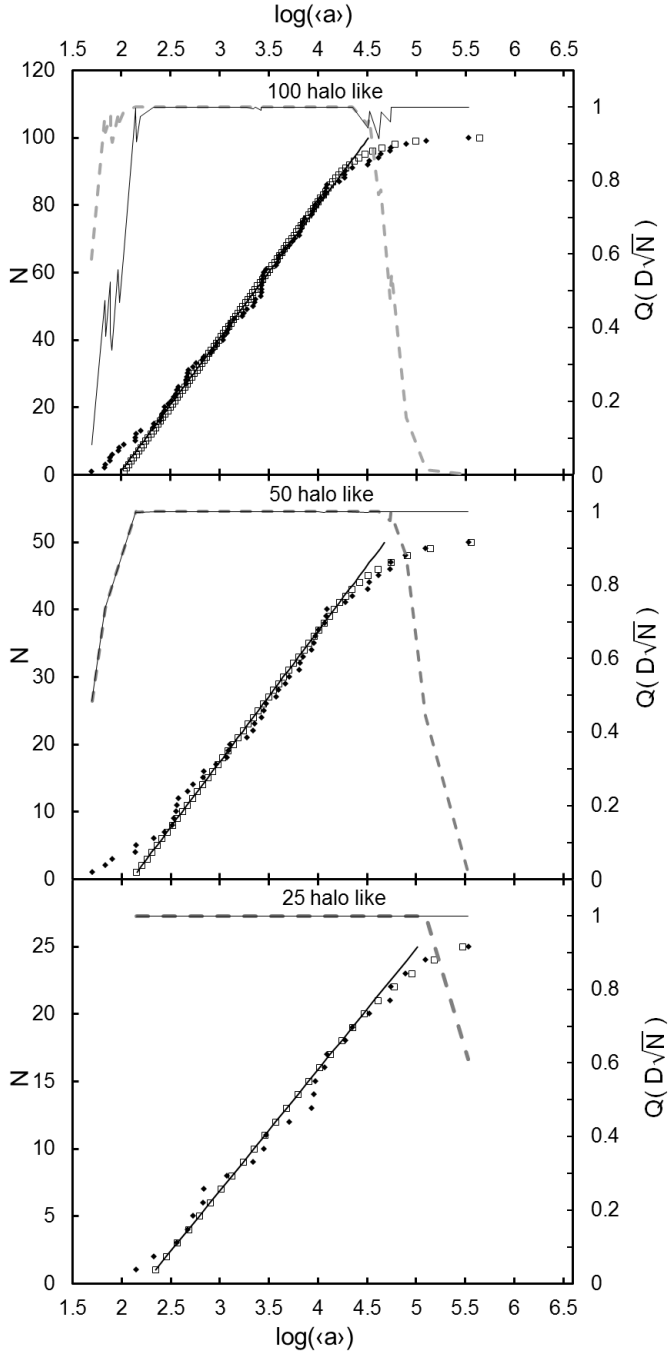}%width=\columnwidth,
\caption{Cumulative distributions of observed semiaxes for three samples of binaries from our catalogue, displayed by diamonds.  The KS test was applied to both a single Oepik-like fit (exponent 1) and to a two-power law fit, with an Oepik like distribution for the unevolved region and a steeper power-law for he evolved region.  The $Q$ indicator for the single power-law fit, plotted on the right hand axis, shows an excellent fit for the unevolved region but drops after a certain value of the semiaxes (dashed lines).  The two power-law fits are able to represent well the entire region, as shown by the empty squares and the $Q$ indicator (full line).}
\label{fig:simple}
\end{figure}

\begin{figure}[!t]
\includegraphics[height=18cm]{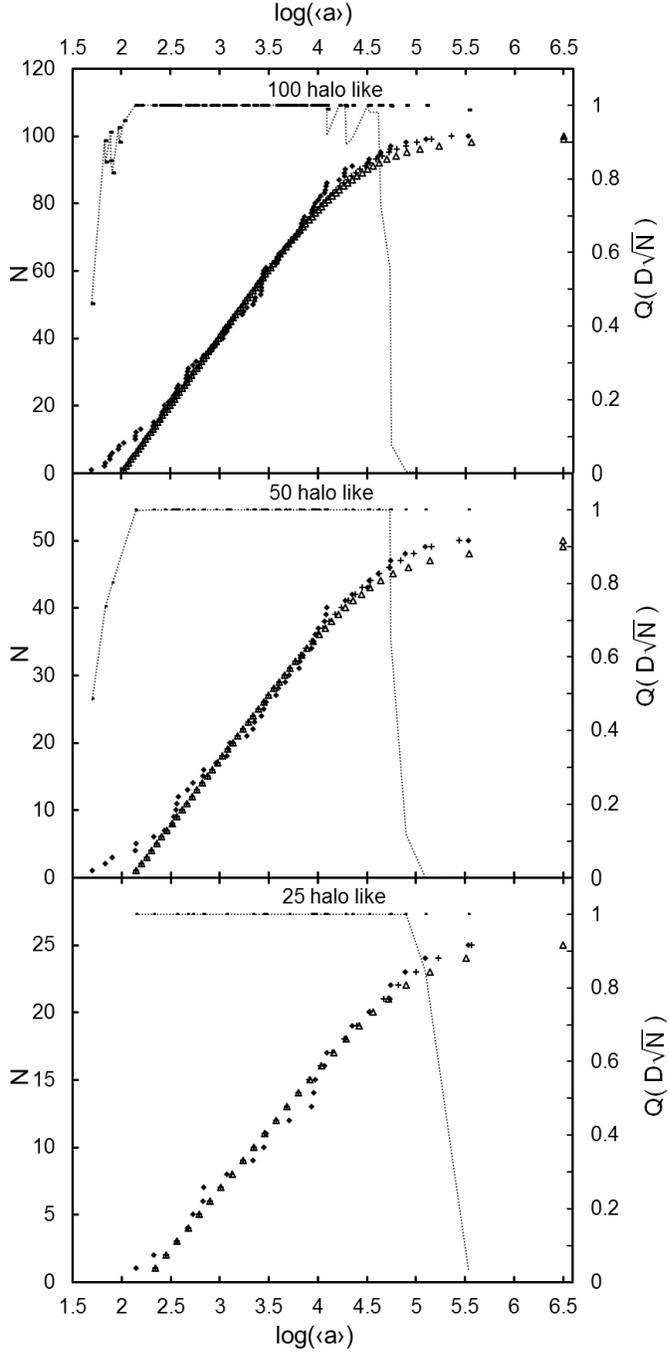}%width=\columnwidth
\caption{KS fits to the evolved binaries.  The thin dotted horizontal line at $Q=1$ (and the small dots prolonguing it) is the $Q$ indicator for the best fit (shown by crosses), which defines $\sigma_0$.  The dropping dotted line is the $Q$ indicator for the still acceptable fit to $2\sigma_0$, shown as triangles.  The diamonds represent the observed samples.  The $2\sigma_0$ fits by this method give maximum perturber masses of 21,11 and $3~M_\sun$ for the 100, 50 and 25 most halo-like stars, respectively.}
\label{fig:simple}
\end{figure}

\begin{figure}[!t]
\includegraphics[width=\columnwidth]{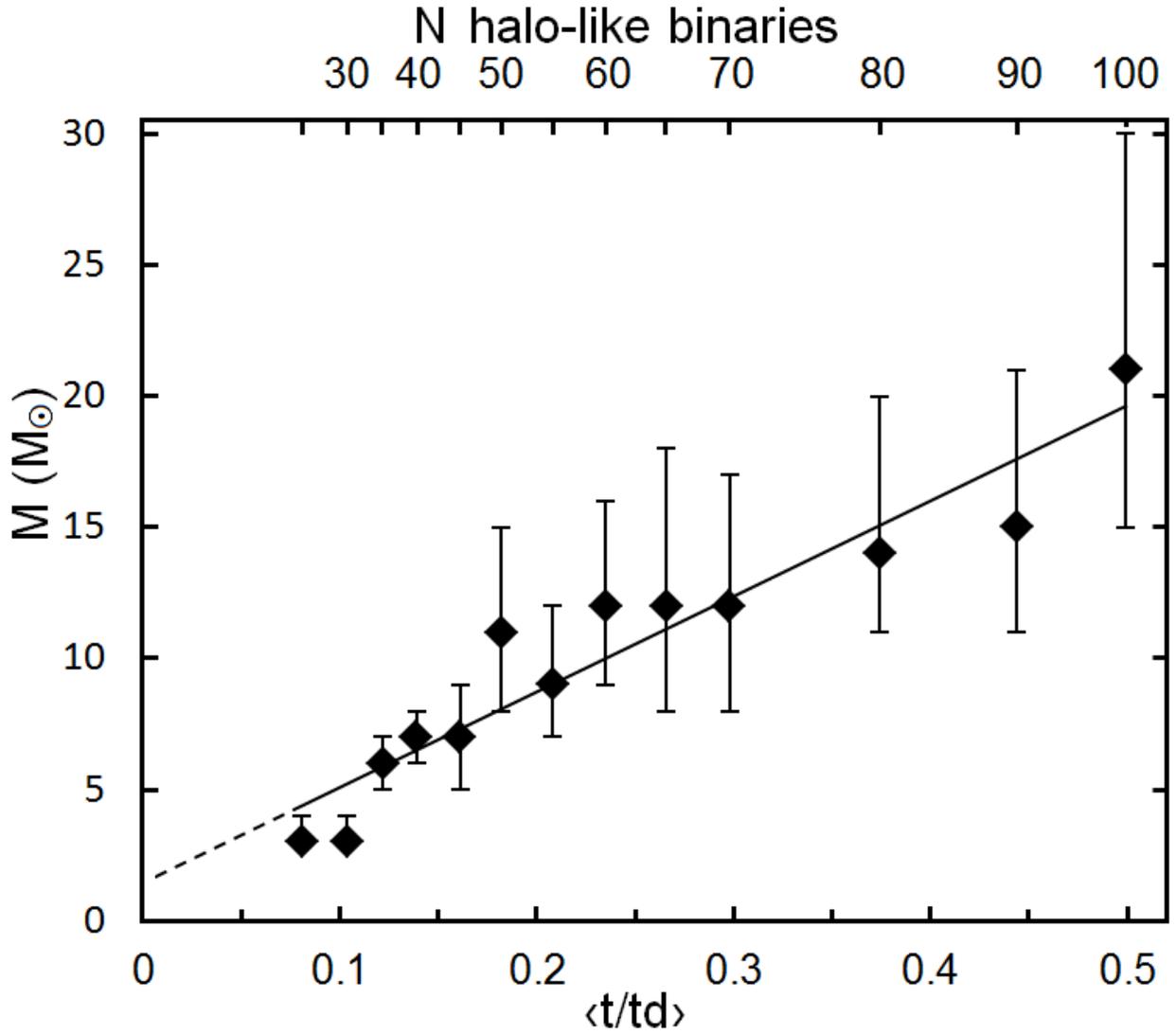}
\caption{Maximum perturber masses for different groups of binaries from our catalog plotted against time spent within the disk. The error bars shown correspond to assumed distance errors of $\pm 20$\%.}
\label{fig:simple}
\end{figure}

\begin{figure}[!t]
\includegraphics[width=\columnwidth]{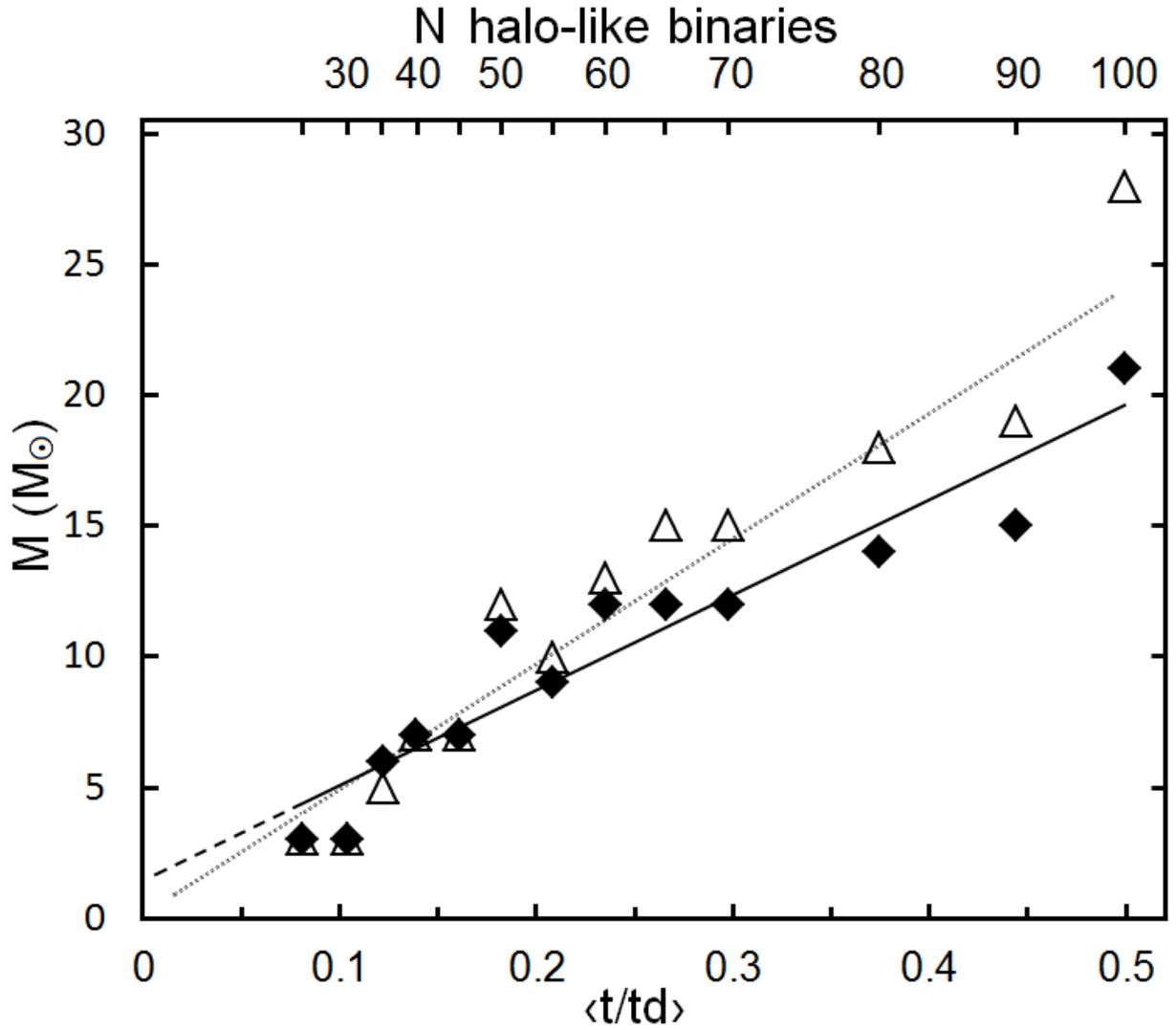}
\caption{Maximum perturber masses for different groups of binaries plotted against time spent within the disk.  The full diamonds correspond to a uniform halo density.  The empty triangles show the effect of taking into account the variable halo density encountered on average by the different groups of binaries along their galactic trajectories}
\label{fig:simple}
\end{figure}

\begin{figure}[!t]
\includegraphics[width=\columnwidth]{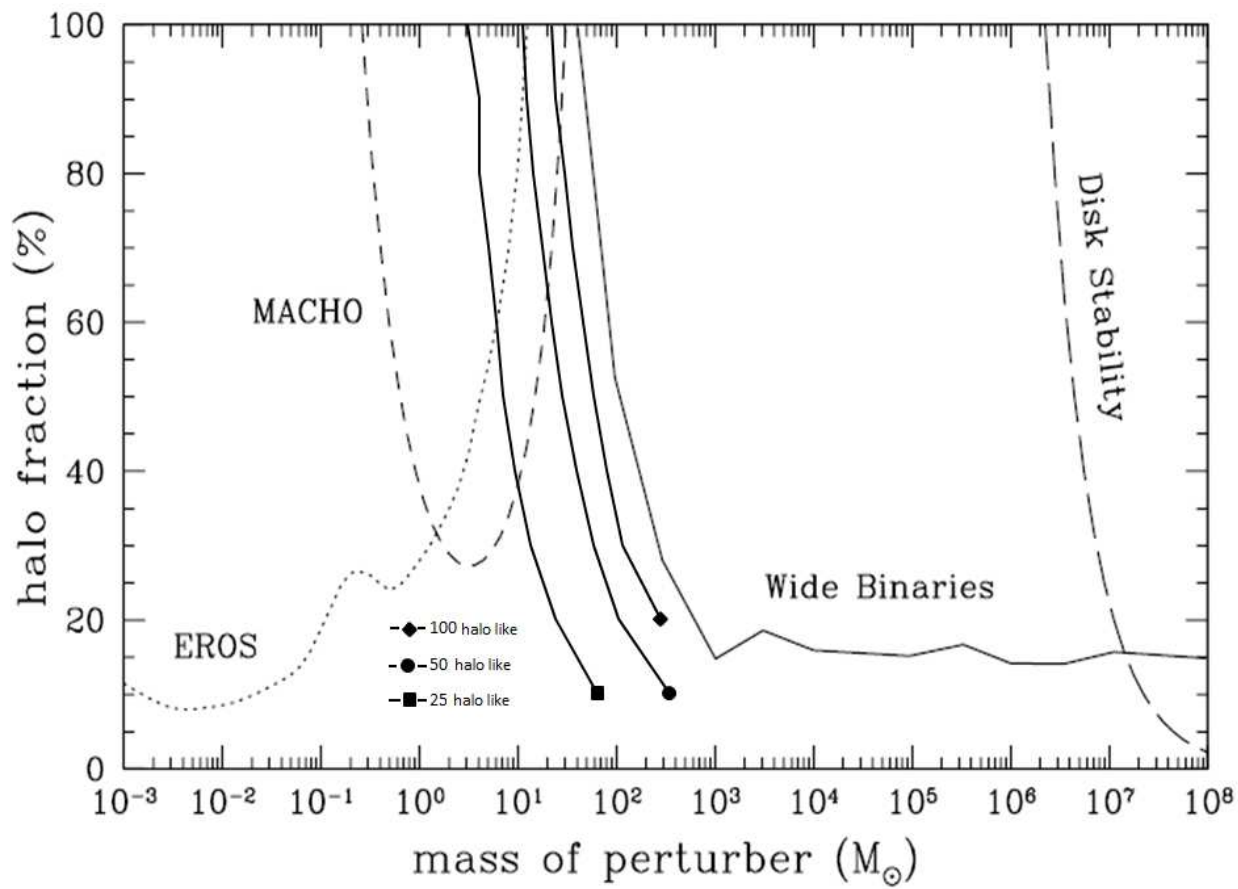}
\caption{Exclusion contours for three different binary samples from our catalog.  For comparison, we plot also the result of Yoo et al., labeled as ``wide binaries''. The contours plotted correspond to the results using unbinned data and the KS criterion $Q$ for goodness of fit (see also Table 1).}
\label{fig:simple}
\end{figure}
\clearpage

\begin{table}
%\begin{center}
\caption{Upper limits for the mass of the Massive Compact Halo Objets (MACHOs).\label{tbl-1}}
\centering
\begin{tabular}{cc cc cc}
\tableline\tableline
\multicolumn{1}{c}{$Sample$\tablenotemark{a}} & $\langle T_{d}/T_{t}\rangle$ & $M(\rho_{0})_b$ & $M(\langle\rho\rangle)_b$& $M(\rho_{0})_{{\rm KS}}$ & $M(\langle\rho\rangle)_{{\rm KS}}$\\
&&$[M_{\sun}]$&$[M_{\sun}]$ &$[M_{\sun}]$&$[M_{\sun}]$\\
%%\multicolumn{1}{c}{$P$\tablenotemark{a}} & $P R_{maj}$ & $P R_{min}$ &
%%\multicolumn{1}{c}{$\Theta$\tablenotemark{b}} \\
\tableline
25&0.08& 12 & 11 & 3 & 3 \\
30&0.10&16&13 & 3 & 3 \\
35&0.12&19&17 & 6 & 5 \\
40&0.14&23&22 & 7 & 7 \\
45&0.16&34&33 & 7 & 7 \\
50&0.18&39&41 & 11 & 12 \\
55&0.21&43&48 & 9 & 10 \\
60&0.24&45&55 & 12 & 13 \\
65&0.27&51&58 & 12 & 15 \\
70&0.30&54&61 & 12 & 15 \\
80&0.37&63&68 & 14 & 18 \\
90&0.44&67&73 & 15 & 19 \\
100&0.50&68&78 & 21 & 28 \\
Yoo\tablenotemark{b}&&80&&31&\\
Quinn\tablenotemark{c}&&105&&32&\\
\hline
\end{tabular}
%% Any table notes must follow the \end{tabular} command.
\tablenotetext{a}{Number of most halo like binaries with computed galactic orbits}
\tablenotetext{b}{90 halo binaries of Chanam\'e and Gould (2004)}
\tablenotetext{c}{89 halo binaries of Quinn et al. (2009)}
%\tablecomments{The $<\rho>$ was estimated asuming ich binary pass $93\% $ of his time in $R_{max}$ and $7\% $ in $R_{min}$}
%\end{center}
\end{table}

\begin{table}
\caption{Effect of distance errors (20\%) on maximum MACHO masses\tablenotemark{*}.\label{tbl-4}}
\begin{center}
\begin{tabular}{cccc}
\tableline\tableline
%\begin{deluxetable}
%\tablewidth{0pc}
%\tablecaption{Effect of distance erros (20\%) on maximum MACHO masses\tablenotemark{*}.\label{tbl-3}}
%\tablehead{
\multicolumn{1}{c}{$Sample$\tablenotemark{a}} & $+20$\% & 0 & $-20$\%\\
\tableline
%\startdata
25 & 3 & 3 & 4 \\
30 & 3 & 3 & 4 \\
35 & 5 & 6 & 7 \\
40 & 6 & 7 & 8 \\
45 & 5 & 7 & 9 \\
50 & 8 & 11 & 15 \\
55 & 7 & 9 & 12 \\
60 & 9 & 12 & 16 \\
65 & 8 & 12 & 18 \\
70 & 8 & 12 & 17 \\
80 & 11 & 14 & 20 \\
90 & 11 & 15 & 21 \\
100 & 15 & 21 & 30 \\
\tableline
\end{tabular}
%\enddata
\tablenotetext{*}{For a constant halo density, adopting Couteau's formula}
\tablenotetext{a}{Number of most halo-like binaries with computed galactic orbits.}
\end{center}
\end{table}

\begin{deluxetable}{lcc cccc cccc r}
\tabletypesize{\scriptsize}
\tablewidth{0pc}
\tablecaption{The 25 most halo like wide binaries}%%\label{tbl-2}}
\tablehead{
%%\colhead{Star} & \colhead{V} &\colhead{b$-$y} & \colhead{m$_1$} &\colhead{c$_1$} & \colhead{ref} &\colhead{T$_{\rm eff}$} & \colhead{log g} &\colhead{v$_{\rm turb}$} & \colhead{[Fe/H]} &\colhead{ref}}
Primary & Secondary & $d$ & $M_V$(p)& $M_V$(s)& s &$\langle{}a\rangle{}$ & $v_p$ & $R_{\rm max}$ & $|z_{\rm max}|$ & $e$ & td/t\\
&&(pc)& & &(")&(AU)&(km/s)&(kpc)&(pc)\\
(1)&(2)&(3)&(4)&(5)&(6)&(7)&(8)&(9) &(10)&(11)&(12)}
\startdata
NLTT 16394\tablenotemark{*}&NLTT 16407\tablenotemark{*}&348&4.5&7.6&698.5&340309&481.5&144.0&88786
&0.90&0.003\\
NLTT 37787&NLTT 37790&193&9.9&6.4&200.9&54283&470&115.2&109015
&0.92&0.003\\
NLTT 4814&NLTT 4817&153&10.7&5.2&24.4&5139&303&12.9&8058&0.98&0.006\\
NLTT 5781&NLTT 5784&306&11.6&7.1&52.1&22320&380.4&44.2&32637
&0.91&0.019\\
NLTT 18775&G090-036B&293&5.4&8.8&1.67&685&515.8&429.1&53116
&0.96&0.020\\
NLTT 18346\tablenotemark{*}&NLTT 18347\tablenotemark{*}&167&5.1&4.3&11.9&2805&290.3&21.6&17469&0.52&0.028\\
NLTT 49562\tablenotemark{*}&G 262-022r\tablenotemark{*}&281&6.7&7.2&29.54&12155&370.5&40.0&27828
&0.98&0.031\\
NLTT 1036&NLTT 1038&290&10.8&6.0&22.5&9135&213&9.8&7363
&0.29&0.049\\
NLTT 32187&VBS 2&80&3.8&7.8&1.9&213&356.1&22.0&9995&0.94&0.055\\
NLTT 41756&VB 6&256&6.3&8.7&1.5&537&515.9&19.9&6958&0.43&0.060\\
NLTT 10536\tablenotemark{*}&NLTT 10548\tablenotemark{*}&214&4.6&9.1&185.7&55410&417.5&9.6&5202&0.12&0.062\\
NLTT 43097\tablenotemark{*}&G 017-027j\tablenotemark{*}&48&6.2&10.5&1170.7&79139&162.7&9.4&5692
&0.69&0.074\\
NLTT 39442&LP 424-27&132&6.4&6.6&677&125073&306.7&15.6&7088&0.87&0.075\\
NLTT 5690&LP 88-69&71&4.1&9.1&22&2198&183.7&8.6&4562&0.51&0.083\\
NLTT 39456\tablenotemark{*}&NLTT 39457\tablenotemark{*}&29&6.7&7.1&300.7&12380&592.2&67.0&6410&0.86&0.088\\
NLTT 16062&BD 34 567&70&6.6&10.4&12&1176&265.5&11.8&6401&0.94&0.103\\
NLTT 9798&LP 470-9&100&3.6&7.6&1&140&104.5&8.7&2366&0.10&0.121\\
NLTT 52786&NLTT 52787&167&5.8&5.3&2&468&162.4&9.7&3623
&0.58&0.123\\
NLTT 16629&NLTT 16631&55&6.4&10.5&434.1&33416&179.6&8.9&2562
&0.29&0.125\\
NLTT 19979&NLTT 19980&344&7.9&10.2&19.6&9439&228.1&15.0&8120
&0.92&0.135\\
NLTT 38195&BD+06 2932B&73&6.1&9.0&3.6&443&267.6&18.1&10301
&0.96&0.137\\
NLTT 51780&G093-027B&136&6.0&8.7&3.52&672&259.8&14.4&2596
&0.45&0.149\\
NLTT 525&NLTT 526&251&4.8&7.4&8.5&2986&361.3&31.6&3122&0.92&0.157\\
NLTT 14005&NLTT 13996&103&5.7&12.6&133&19173&101.9&8.7&2000&0.20&0.157\\
NLTT 17770&LP 359-227&180&4.1&10.4&34&8565&262.7&11.7&8661&0.96&0.158\\
\enddata
\tablenotetext{*}{wide binaries with concordant radial velocities.}
%%\end{center}
\end{deluxetable}

\begin{deluxetable}{crr}
\tablewidth{0pc}
\tablecaption{Comparison of upper limits to the MACHO masses using two estimates for $\langle a\rangle$, for a constant halo density.\label{tbl-3}}
\tablehead{
& Couteau & Bartkevicius\\
\multicolumn{1}{c}{$Sample$\tablenotemark{a}} & $M(\rho_{0})$ & $M(\rho_{0})$}
\startdata
25 & 3 & 4\\
30 & 3 & 4\\
35 & 6 & 6\\
40 & 7 & 8\\
45 & 7 & 8\\
50 & 11 & 12\\
55 & 9 & 9\\
60 & 12 & 13\\
65 & 12 & 14\\
70 & 12 & 14\\
\enddata
\tablenotetext{a}{Number of most halo-like binaries with computed galactic orbits.}
\end{deluxetable}

\end{document}